\documentclass[11pt]{article}
\usepackage{moriond2000}
\usepackage{epsfig}
\newcommand{\Journal}[4]
        {{#1} {\bf #2}, #3 (#4).}
\newcommand{\SubJournal}[3]
        {{#1} #2 (#3).}

\newcommand{\singlefigureheight}
	{65mm}

\newcommand{\la}
	{<}
\newcommand{\ga}
	{>}

\newcommand{\sect}[1]
	{Section~\ref{section:#1}}
\newcommand{\fig}[1]
        {Fig.~\ref{figure:#1}}

\newcommand{\ie}
	{{\em i.e.}}
\newcommand{\eg}
	{{\em e.g.}}

\newcommand{\ApJ}
	{ApJ}
\newcommand{\ARAA}
        {ARA\&A}
\newcommand{\MNRAS}
        {MNRAS}

\newcommand{\AJ}
        {AJ}
\newcommand{\PhilTransA}
	{Phil.\ Trans.\ Royal Ast.\ Soc.\ A}

\begin{document}
\vspace*{4cm}
\title{THE STATISTICS OF WIDE-SEPARATION LENSES}

\author{ D.J.\ MORTLOCK$^{1,2,3}$}

\address{$^{1}$Astrophysics Group, Cavendish Laboratory, Madingley Road,
        Cambridge CB3 0HE, U.K. \\
        $^{2}$Institute of Astronomy, Madingley Road, Cambridge
        CB3 0HA, U.K. \\
	$^{3}$Department of Physics, The University of Melbourne, Parkville,
        Victoria 3052, Australia}

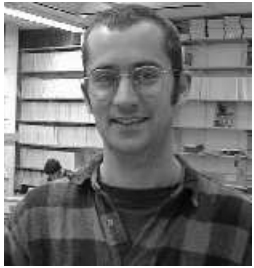
\maketitle\abstracts{The probability that high-redshift sources are 
gravitationally-lensed with large image separations (\ie, greater than can 
be produced by galactic deflectors) is determined by the cosmological 
population of group- and cluster-sized halos. Thus the observed frequency of 
wide-separation lensed quasars can be used to constrain not only 
the halo distribution, but also a number of cosmological parameters.
A calculation of the optical depth due to collapsed, isothermal halos
is a useful guide to the lens statistics, and illustrates that
the number of wide-separation lenses is a sensitive probe of 
the mean density of the universe
and
the present day density variance
whilst being nearly independent 
of the cosmological constant.}

\section{Introduction}
\label{section:intro}

Gravitational lensing is a direct probe of the geometry of the 
universe, and so observations of lensing place constraints on 
both the distribution of mass and the
underlying cosmological model. 
A large variety of lensing experiments are possible:
weak shear surveys;
searches for microlensing by compact objects; 
observations of giant arcs in clusters;
measurement of Shapiro delay in binary pulsars;
surveys for multiply-imaged qusasrs; and so on. 
The focus here is on the 
statistics of wide-separation lensed quasars (which are produced
by group- and cluster-mass objects). 
Using a simple model of the halo population
(\sect{deflectors}),
the optical depth to multiple imaging and the expected distribution
of image separations can be calculated (\sect{calc}).
The results obtained are independent of the source population,
which then implies that a more sophisticated analysis is 
required to constrain model parameters from the observed frequency
wide-separation lenses~\cite{ko95}$^{,}$~\cite{mo00}.

\section{Deflector population}
\label{section:deflectors}

\begin{figure}
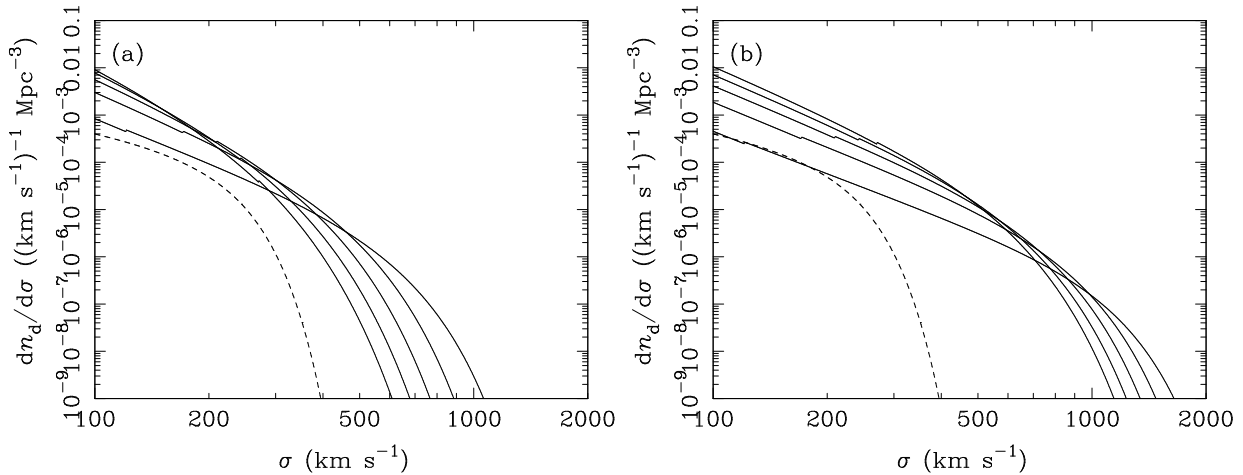

\includegraphics{dndsig_0.5.ps}
\includegraphics{dndsig_1.0.ps}
\vspace{\singlefigureheight}
\caption{The evolution of the halo population in
Press-Schechter theory.
A standard CDM model (with $\Omega_{\rm m_0} = 1$
and $\Omega_{\Lambda_0} = 0$) is assumed,
with $\Delta_8 = 0.5$ in (a) and $\Delta_8 = 1.0$ in (b).
For each model the five solid lines show the population
at redshifts of 0, 1, 2, 3 and 4;
the number of high-mass halos decreases with redshift
(or, equivalently, increases with time).
The dashed lines show the local galaxy population, assuming 
standard magnitude-velocity dispersion relationships.} 
\label{figure:dndsig_ps}
\end{figure}

The population of cluster-sized halos is well approximated
by the Press-Schechter~\cite{pr74} mass function,
which depends on the
cosmological model~\footnote{This is specified by 
the present day normalised matter density,
$\Omega_{\rm m_0}$,
the similarly normalised cosmological constant,
$\Omega_{\Lambda_0}$,
and Hubble's constant (although its value is unimportant in this 
calculation).} and the matter power spectrum. 
An approximate cold dark matter (CDM) power spectrum~\cite{ef92}
with a spectral slope of $n = 1$ is assumed, and only the normalisation
is allowed to vary.
The scale of the density fluctuations are normalised to match 
$\Delta_8$, 
the present day variance in spheres of 
8 Mpc radius, and so is linked to a (co-moving) scale, rather than
a mass.
The clusters themselves are modelled as singular isothermal spheres,
which are characterised by their line-of-sight velocity dispersion,
$\sigma$, rather than their mass, $M$.
General arguments imply that 
$M \propto \sigma^{3}$, but this conversion is somewhat 
ambiguous, and is an important source of systematic error in the 
calculation of lensing cross-sections.
The resultant halo population (\fig{dndsig_ps}) is 
most stronly dependent on $\Delta_8$, and can be made consistent 
with galaxy counts if the Press-Schecter form is assumed only
for $\sigma \ga 200$ km s$^{-1}$.

\section{Lensing optical depth}
\label{section:calc}

Given the population of deflectors and a lens model, it is reasonably
straightforward to calculate the optical depth, $\tau(z_{\rm s})$, 
to multiple imaging~\cite{tu84}$^{,}$~\cite{ko95}. 
Whilst $\tau$ is too crude an estimate of the lensing probability
to meaningfully constrain model parameters,
it is useful in an illustrative sense, particularly if the 
source-dependent aspects of the full probability (\eg, the
magnification bias) can be factored out of the integral over
the deflector population. This is the case for the isothermal
sphere, and, in addition, the image separation, $\Delta \theta$, is 
completely determined by $z_{\rm d}$, $z_{\rm s}$ and $\sigma$,
and so image separation cut-offs
(\ie, $\Delta \theta_{\rm min}$ and $\Delta \theta_{\rm max}$) can be
included in $\tau$.
Thus the optical depth can 
be thought of as the fraction of the sphere at redshift $z_{\rm s}$ 
inside the Einstein radius ($\theta_{\rm E}$) of any cluster 
for which $\Delta \theta_{\rm min} \leq 2 \theta_{\rm E} \leq 
\Delta \theta_{\rm max}$. 
Choosing $\Delta \theta_{\rm min} \simeq 3$ arcsec 
removes the galactic lenses from the calculation (their population
being poorly approximated by the naive Press-Schechter form); 
the value $\Delta \theta_{\rm max}$ is determined by the 
breadth of the companion search. 

\begin{figure}
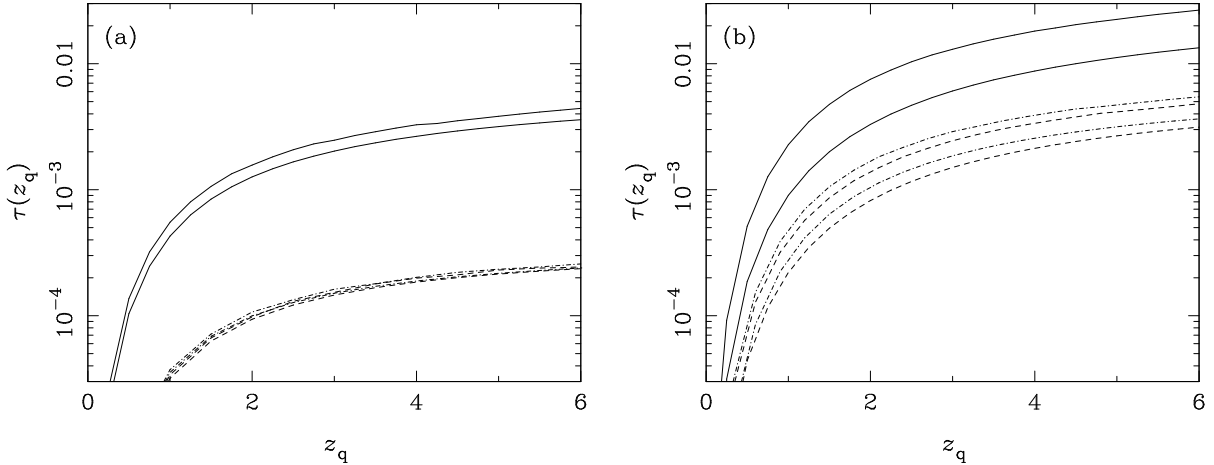

\includegraphics{tau_0.5.ps}
\includegraphics{tau_1.0.ps}
\vspace{\singlefigureheight}
\caption{The gravitational lensing optical depth due to a
cosmological population of isothermal
halos as a function of source redshift. Results are shown for 
several cosmological models: 
$\Omega_{\rm m_0} = 1$ and $\Omega_{\Lambda_0} = 0$ (solid lines);
$\Omega_{\rm m_0} = 0.3$ and $\Omega_{\Lambda_0} = 0$ (dashed lines);
and 
$\Omega_{\rm m_0} = 0.3$ and $\Omega_{\Lambda_0} = 0.7$ (dot-dashed lines),
with $\Delta_8 = 0.5$ in (a) and 
$\Delta_8 = 1.0$ in (b).
The effects of image separation cut-offs are also illustrated:
in each case the lower lines are for
$\Delta \theta_{\rm min} = 3$ arcsec and 
$\Delta \theta_{\rm max} = 10$ arcsec
and the upper lines are for
$\Delta \theta_{\rm min} = 3$ arcsec and
and $\Delta \theta_{\rm max} = \infty$.}
\label{figure:tau}
\end{figure}

The optical depth is shown as a function of 
source redshift in \fig{tau}. The standard increase of 
$\tau$ with $z_{\rm s}$ is apparent, and comparing \fig{tau} (a) and (b)
shows the expected dependence on $\Delta_8$. 
More interesting is the variation with cosmological model and 
$\Delta \theta_{\rm max}$. 
The optical depth to lensing by galaxies is primarily dependent on 
the cosmological constant as the differential volume element is 
so much larger in high-$\Omega_{\Lambda_0}$ models.
This effect is present here, but it is not dominant, for two reasons.
Firstly, clusters form earlier in low-density cosmologies -- 
most large-scale structure formed before 
$z \simeq \Omega_{\rm m_0}^{-1} - 1$ (if $\Omega_{\Lambda_0} = 0$)
or, more recently, before $z \simeq \Omega_{\rm m_0}^{-1/3} - 1$ 
(in flat models)~\cite{ca92}.
Thus the lensing probability is greater if $\Omega_{\rm m_0}$ is small,
there being more high-redshift 
collapsed deflectors along a given line-of-sight.
For a fixed $\Omega_{\rm m_0}$, however, increasing 
the cosmological constant reduces the number of halos 
with $\Omega_{\rm m_0}^{-1/3} - 1 \la z_{\rm d} \la \Omega_{\rm m_0}^{-1} - 1$, 
somewhat offsetting the usual
increase of $\tau$ with $\Omega_{\Lambda_0}$.
Secondly, the mass of a cluster that has collapsed from 
a given co-moving scale (\eg, 8 Mpc) is proportional to 
$\Omega_{\rm m_0}$. For the isothermal sphere model the
lensing cross-section scales as 
$\sigma^4 \propto M^{4/3} \propto \Omega_{\rm m_0}^{4/3}$. 
It is because of this strong dependence that
standard CDM models with $\Omega_{\rm m_0} = 1$ 
are so inconsistent with the low number of wide separation
lenses~\cite{mo00}$^{,}$~\cite{ko95}$^{,}$~\cite{na88}.

\begin{figure}
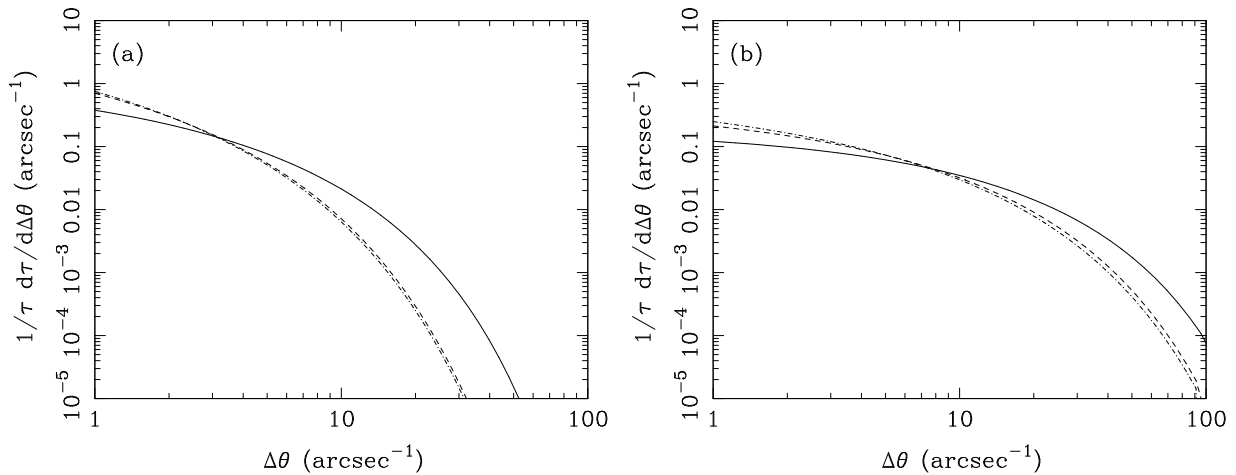

\includegraphics{theta_0.5.ps}
\includegraphics{theta_1.0.ps}
\vspace{\singlefigureheight}
\caption{The normalised distribution of image separations produced by a
cosmological population of isothermal halos, for a source 
at $z_{\rm q} = 3$.
Results are shown for
several cosmological models:
$\Omega_{\rm m_0} = 1$ and $\Omega_{\Lambda_0} = 0$ (solid lines);
$\Omega_{\rm m_0} = 0.3$ and $\Omega_{\Lambda_0} = 0$ (dashed lines);
and
$\Omega_{\rm m_0} = 0.3$ and $\Omega_{\Lambda_0} = 0.7$ (dot-dashed lines),
with $\Delta_8 = 0.5$ in (a) and
$\Delta_8 = 1.0$ in (b).}
\label{figure:theta}
\end{figure}

The expected distribution of image separations can be estimated by
computing ${\rm d}\tau / {\rm d}\Delta \theta$, which is illustrated 
in \fig{theta}.
As shown, these distributions are normalised to unity, but 
the vertical scaling can be estimated from the known
wide-separation pairs~\cite{na88}.
The immediate implication of this is that the fall-off with
increasing $\Delta \theta$ is far too shallow to be consistent
with the complete absence of any (confirmed) lenses
with $\Delta \theta \ga 10$ arcsec~\cite{ko95}.
One inference that could be drawn is that the mass profiles 
of clusters are shallower than isothermal~\cite{ma00},
but it is difficult to reconcile such models
(including the Navarro, Frenk \& White~\cite{na97} profile)
with observations of arcs and arclets~\cite{wi99}.

\section{Conclusions}
\label{section:concs}

The statistics of wide-separation lenses are a useful cosmological 
probe. The lensing probability 
is sensitive to the geometry of the universe, and the 
population of collapsed halos at intermediate redshifts whilst
being independent of the complexities of galaxy formation. 
As large-scale structure is dependent on the linear growth of 
perturbations (and also because the mass in a co-moving sphere of 
a given size is proportional to the density parameter) the
likelihood of lensing is most sensitive to $\Omega_{\rm m_0}$ and
$\Delta_8$. The 
Large Bright Quasar Survey~\cite{he95} 
is devoid of wide-separation 
lenses~\cite{he98}$^{,}$~\cite{mo99}
which implies that 
$\Omega_{\rm m_0} \la 0.3$ (assuming $\Delta_8 \ga 0.4$)
or that
$\Delta_8 \la 0.6$ (assuming $\Omega_{\rm m_0} \la 0.1$)
with 99 per cent confidence~\cite{mo00}.
In the future such analyses could be extended to the 
2 degree Field quasar survey~\cite{bo99} (with $\sim 3 \times 10^4$ quasars)
and the Sloan Digital Sky Survey~\cite{sz98} (which will contain 
$\sim 10^5$ sources). These samples will be large enough to constrain
$\Omega_{\rm m_0}$ and $\Delta_8$ to within several per cent 
from the number of wide-separation lenses alone.

\section*{Acknowledgments}
Much of this research was done whilst supported by an 
Australian Postgraduate Award.

\end{document}